\documentclass[journal]{IEEEtran}

\usepackage{cite,graphicx,amssymb,amsmath,color,textcomp}
\usepackage{subfigure}
\usepackage{mathrsfs}
\usepackage{supertabular}
\usepackage{tabularx}
\usepackage{longtable,stfloats,enumerate}
\usepackage{multicol,multirow,epstopdf}

\newtheorem{remark}{\underline{Remark}}

\begin{document}

\title{ Interference-Free Transceiver Design and Signal Detection for Ambient Backscatter Communication Systems over Frequency-Selective Channels}
\author{Chong Zhang, Gongpu Wang, Panagiotis D. Diamantoulakis, ~\IEEEmembership{Member, IEEE}, Feifei Gao, ~\IEEEmembership{Senior Member, IEEE}, and George K. Karagiannidis, ~\IEEEmembership{Fellow, IEEE}\\

\thanks{C. Zhang is with Beijing Intelligent Logistics System Collaborative Innovation Center and  School of Electronic and Information Engineering, Beijing Jiaotong University, Beijing 100044, P. R. China  (email: 15212111@bjtu.edu.cn).}
\thanks{G. Wang is with Beijing Key Lab of Transportation Data Analysis and Mining, School of Computer and Information Technology, Beijing Jiaotong University, Beijing 100044, P. R. China (e-mails: {gpwang}@bjtu.edu.cn).}
\thanks{F. Gao is with Department of Automation, Tsinghua University, State Key Lab of Intelligent Technologies and Systems, Tsinghua National Laboratory for Information Science and Technology (TNList) Beijing 100084, P. R. China (email: feifeigao@ieee.org).}
\thanks{P. D. Diamantoulakis and G. K. Karagiannidis are with the Department of Electrical and Computer Engineering, Aristotle University of Thessaloniki, Thessaloniki 54124, Greece (e-mails: padiaman, geokarag@auth.gr).}
\thanks{ Corresponding author: Gongpu Wang (gpwang@bjtu.edu.cn).}

}
\maketitle

\begin{abstract}
In this letter, we study the  ambient backscatter communication systems
over frequency-selective channels.
Specifically, we propose an interference-free transceiver design
to   facilitate    signal detection    at  the reader.
Our  design    utilizes
the cyclic prefix (CP)
of orthogonal frequency-division multiplexing (OFDM)  source
symbols, which can cancel the signal interference
and thus   enhance  the detection accuracy  at the reader.
Meanwhile, our design  results in  no interference  on  the
existing OFDM communication systems.
We also suggest a maximum likelihood (ML) detector for
the reader and derive two detection thresholds.
Simulations are then  provided to corroborate our  proposed studies.
\end{abstract}

\begin{IEEEkeywords}
Ambient backscatter,  frequency-selective channels,
signal detection,  wireless communications.
\end{IEEEkeywords}

\section{Introduction}
As a newborn   green   technology   for   the Internet of Things (IoT),
ambient backscatter \cite{Lu_2018_survey},
has attracted extensive attention \cite{Wang_2016ITComm, Qian_2017_TWC, Yang_2018_TWC, Yang_2018_Internet}.
This novel technology
utilizes ambient radio frequency (RF) signals to implement the
backscatter communications of low data-rate devices such as
tags or sensors, and is able to free them from batteries.

\begin{figure}[t]
\centering
\includegraphics[height=53mm,width=75mm]{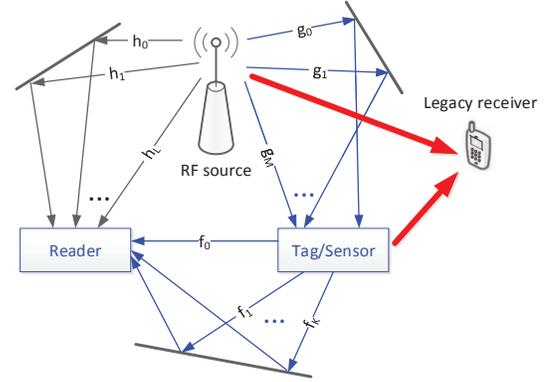}
\caption{System model. }
\label{fig:System_Model} \vspace{-5mm}
\end{figure}

A typical ambient backscatter communication system
includes three components:  a RF source, a tag (or a sensor),  and a reader,
as shown in Fig. \ref{fig:System_Model}.
The communication  process   between the tag and the reader  mainly contains two steps:
first, the tag  harvests energy from the  signals  of  the RF  source;
second, the tag  modulates its binary information
  onto the received RF signals and  then backscatters them  to   the reader.

In the open literature, almost all studies
about ambient backscatter communication
are based on the assumption of   flat-fading channels.
Nevertheless,
the frequency-selective channels
widely exist
in numerous application scenarios.
For   ambient backscatter communication systems,
  frequency-selective channels
  could  lead to multiple copies of backscattered  signals at the reader,
   together with   multiple copies of   source signals.
  Consequently,  it is   one   challenging  problem  for  the reader
  to realize recovery of tag binary information.

In this letter, we investigate the
ambient backscatter communication systems
over frequency-selective channels  and
  propose an interference-free   transceiver design
to cope with  the signal detection challenge  at the reader.
In our proposed  design,
the cyclic prefix (CP) structure
of orthogonal frequency-division multiplexing (OFDM) source
symbols is exploited, which can  benefit signal detection at the reader
via cancelling the signal interference.
Meanwhile, different from the transceiver design in \cite{Yang_2018_TWC}, our design leads
to no interference to the legacy receivers.
Next, a maximum likelihood (ML) detector
 is  proposed  and  two  detection thresholds
   are derived.
   Our  simulation results  show that
our transceiver design for the frequency-selective channels
is efficient and achieves low bit error rate (BER) due to
interference cancellation.

The rest of this paper is organized as follows:
Section \ref{sec:System Model} formulates the model of the ambient backscatter communication
system over frequency-selective channels.
Section \ref{sec:Transceiver Design}
proposes the transceiver design
and Section \ref{sec:Signal Detection at the Reader} derives the ML detector
with two detection thresholds.
Section \ref{sec:Simulation Results} provides
the simulation results
and finally Section \ref{sec:Conclusion} summarizes this letter.

\section{System Model}
\label{sec:System Model}
Consider an ambient backscatter communication system
 over frequency-selective channels
   in Fig. \ref{fig:System_Model}.
 The   multi-path    channels between the
  RF source and reader,
  the  RF source and
    tag,
   the reader and  tag
  are denoted by
  $h_l \,\,(l=0, 1, \cdots, L)$, $g_m \,\,(m=0, 1, \cdots, M)$, $f_k \,\,(k=0, 1, \cdots, K)$,
  separately.
Both the reader and the legacy receiver can receive signals from
the RF source and the tag over frequency-selective channels.

The   signal transmitted by the RF source is $s(n)$
with a zero-mean and a variance of $P_s$ and $s(n) \sim \mathcal{CN}(0,P_s)$.
Due to the multi-path channels $g(m)\,\, (m=0, 1, \cdots, M)$,  the signal
  arriving    at  the tag antenna
can be  given as
\begin{align}
x(n)=\sum_{m=0}^{M} g_m s(n-m).
\end{align}

The tag  next
modulates its own binary signal $B(n)$
onto the received signal $x(n)$
to   communicate with
 the reader via
backscattering  $x(n)$  or not.
Specifically,   the   tag   changes  its antenna impedance
   to   reflect    $x(n)$     to the reader
so as to indicate  $B(n)=1$;
and  when  indicating   $B(n)=0$,
  the tag switches   the impedance to a certain value
   so that no signal
 can be reflected.
 Suppose that the tag information
 $B(n)=0$   and  $B(n)=1$ are equally probable.

Finally, the received signal at the reader can be expressed as
\begin{align}
y(n)=   \sum_{l=0}^{L} h_l s(n-l) + \eta \sum_{k=0}^{K} f_k B(n-k) x(n-k) +w(n),
\end{align}
where  $\eta$ represents the complex  attenuation inside the tag,
$w(n)$  denotes  the additive white Gaussian
noise (AWGN)  and we assume $w(n) \sim \mathcal{CN}(0,N_w)$.



\section{Interference-free Transceiver Design}
\label{sec:Transceiver Design}
In this section, we describe an interference-free transceiver design,
whose implementation   mainly   consists of three crucial aspects:
the   tag  signal   design,
the signal interference cancelling    method,
and the discrete   Fourier transformation (DFT) operation.

\subsection{Tag Signal Design}

\begin{figure}[t]
\centering
\includegraphics[height=45mm,width=90mm]{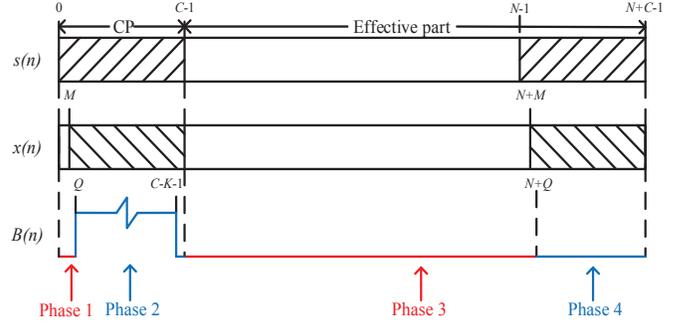}
\caption{The signal structure   at the tag. }
\label{fig:Signal_Structure} \vspace{-3mm}
\end{figure}

With the assumption that the RF source emits OFDM symbols,
the signal structure in one OFDM symbol period at the tag is presented in Fig. \ref{fig:Signal_Structure}.
We set $C$ and $N$ as the lengths of the CP and the effective part of OFDM symbol, respectively.
The parameter $Q$ is defined as $Q={\rm max}\lbrace L, M, K\rbrace$.

Obviously, both $s(n)$ and $x(n)$
have repeating sequences, even if the signal $x(n)$ experiences the
multi-path channels $g(m)\,\, (m=0, 1, \cdots, M)$.
Besides, we divide one OFDM symbol period into four
phases for the designed tag signal $B(n)$.
In Phase 1, Phase 3, and Phase 4, no received signal $x(n)$  will be reflected, i.e., $B(n)=0$.
In this case,  the signals arriving at the reader directly come from the RF source.
However, in Phase 2,  the tag modulates its binary data onto the signal $x(n)$ from
the time $Q$ to the time $C-K-1$ while no signal is backscattered to the reader
in the rest of the Phase 2.
By exploiting the signals arriving at the reader in Phase 2 and Phase 4,
we can cancel the signal interference, which will be presented in
the next subsection.

\begin{remark}
It can be checked from Fig. \ref{fig:Signal_Structure} that the signal structure of $B(n)$
solely effects the samples in the CP of the OFDM symbol.
Since the CP will be removed at the legacy receiver,
this transceiver design at the tag will
lead to no interference to the legacy receivers.
\end{remark}

\subsection{Signal Interference Cancelling Method}
\label{subsec:Signal_interference_cancelling}
Denote the
received signals at the reader
in Phase 2 and Phase 4
as $y_{1}(n)\,\,\,(n=Q, \cdots, C-1)$ and $y_{2}(n)\,\,\,(n=N+Q, \cdots, N+C-1)$, respectively.
We can obtain
\begin{align}
y_{1}(n)=&    \sum_{l=0}^{L} h_l s(n-l) + \eta \sum_{k=0}^{K} f_k B(n-k) x(n-k)   \label{eq:y1_received_signal}\nonumber\\
             &       \qquad\qquad\qquad\qquad\qquad\qquad\qquad +w_1(n),   \\
y_{2}(n)=&    \sum_{l=0}^{L} h_l s(n-l) +w_2(n),  \label{eq:y2_received_signal}
\end{align}
where $w_1(n)$ and $w_2(n)$ are both AWGN. Assume that $w_1(n) \sim \mathcal{CN}(0,N_w)$ and
$w_2(n) \sim \mathcal{CN}(0,N_w)$.


Apparently, the term
$\sum_{l=0}^{L} h_l s(n-l)$ in (\ref{eq:y1_received_signal}) and (\ref{eq:y2_received_signal})
 carries no tag binary information
 and thus is the interference
for the tag signal recovery at the reader,
 which can be cancelled for
 further improvement of detection accuracy.

Signal interference cancelling is implemented via subtracting $y_2(n)$ from $y_1(n)$,
thus the received signals can be written as
\begin{align}
z(n)= & y_{1}(n+Q)-y_{2}(n+N+Q)\nonumber \\
    = & \eta \sum_{k=0}^{K} f_k B(n-k) x(n-k) +w_1(n) -w_2(n),\nonumber \\
    = & \eta \sum_{k=0}^{K} f_k B(n-k) x(n-k) +w_e(n),
\label{eq:Receive_signals_self_interference_elimitation}
\end{align}
where $n=0, 1, \cdots, C-Q-1$ and $w_e(n) \sim \mathcal{CN}(0,2N_w)$.

\subsection{DFT Operation}
Assume $T=C-Q-1$ and $R=C-Q-K-1$. After signal interference cancelling at the reader,
let us construct the signal vector $\textbf{z}$ as
\begin{align}
\textbf{z}= & [z(0)+z(R+1), \cdots, z(T-R-1)+z(T),  \nonumber\\
&   \qquad\qquad\,\,\,\,\, z(T-R-1), \cdots, z(R-1), z(R)]^{\rm T}.
\end{align}

Define
\begin{align}
\textbf{b}= & [B(0), B(1), \cdots, B(n),  \cdots, B(R)],\\
\textbf{x}= & [x(0), x(1), \cdots, x(n),  \cdots, x(R)]^{\rm T},\\
\textbf{w}= & [w_e(0)+w_e(R+1), \cdots, w_e(T-R-1)+w_e(T), \nonumber\\
 & \qquad\qquad\;\, w_e(T-R-1), \cdots, w_e(R-1), w_e(R)]^{\rm T}.
\end{align}

Denote $\textbf{F}$ as the $(R+1)\times (R+1)$  DFT matrix
with the  $(p,q)$th element $\textbf{F}_{pq}=\exp (-j2\pi pq/(R+1))$.
Let us consider a Toeplitz matrix $\textbf{T}$, which possesses the first row of $\textbf{t}_{\textbf{r}}=[f_0, 0, \cdots, 0,  f_K, f_{K-1}, \cdots, f_k,  \cdots, f_1]$ and the first column of $\textbf{t}_{\textbf{c}}=[f_0, f_1, \cdots, f_k,  \cdots, f_K, 0, \cdots, 0]^{\rm T}$.

Consequently, we   reconstruct the signal vector $\textbf{z}$ based on DFT as, i.e., DFT outputs $\tilde{\textbf{z}}$
\begin{align}
\tilde{\textbf{z}}= & \,\textbf{F}\textbf{z}  \nonumber \\
= & \,\eta \textbf{F}  \textbf{T} \cdot {\rm diag}[\textbf{b}] \cdot \textbf{x} + \textbf{F}\textbf{w} \nonumber  \\
= & \,\eta \cdot {\rm diag}\left[\tilde{\textbf{f}}^{\rm T}\right] \cdot {\rm diag}[\textbf{b}] \cdot \tilde{\textbf{x}} + \tilde{\textbf{w}},
\end{align}
where
\begin{align}
\tilde{\textbf{z}}=\, & [\tilde{z}(0), \tilde{z}(1), \cdots, \tilde{z}(n),  \cdots, \tilde{z}(R)]^{\rm T}=\textbf{F}\textbf{z},\\
\tilde{\textbf{x}}=\, & [\tilde{x}(0), \tilde{x}(1), \cdots, \tilde{x}(n),  \cdots, \tilde{x}(R)]^{\rm T}=\textbf{F}\textbf{x},\\
\tilde{\textbf{w}}=\, & [\tilde{w}_e(0), \tilde{w}_e(1), \cdots, \tilde{w}_e(n),  \cdots, \tilde{w}_e(R)]^{\rm T}=\textbf{F}\textbf{w},\\
\tilde{\textbf{f}}=\, & [\tilde{f}_0, \tilde{f}_1, \cdots, \tilde{f}_n,  \cdots, \tilde{f}_R]^{\rm T}=\textbf{F}\textbf{t}_{\textbf{r}}^{\rm T}.
\end{align}

According to the central limit theorem (CLT), we can have $\tilde{x}(n)\sim \mathcal{N}(0, P_x)$, $\tilde{w}_e(n)\sim \mathcal{N}(0, P_w)$ and $\tilde{f}_n\sim \mathcal{N}(0, P_f)$,
 where
\begin{align}
P_x=& (R+1)P_s\sum_{m=0}^{M}|g_m|^2,\\
P_w=& 2(T+1)N_w,\\
P_f=& \sum_{k=0}^{K}|f_k|^2.
\end{align}

\section{Signal  Detection at the Reader}
\label{sec:Signal  Detection at the Reader}
In this section, the ML detector
together with two detection thresholds:
the optimal threshold and the equiprobable error threshold,
is derived.
Moreover, the bit error rate (BER) expression
is obtained for performance analysis.

\subsection{ML Detector}
Due to the lower data-rate of the  tag signal $B(n)$ than that of the signal $\tilde{z}(n)$, we suppose the signal $B(n)$ remains
equivalent within $W$ samples of $\tilde{z}(n)$.
Let us construct the test statistic as
\begin{align}
\Gamma_t=\frac{1}{W}\sum_{n=(t-1)W+1}^{tW} \left|\tilde{z}(n)\right|^{2},
\end{align}
where $t=1, 2, \cdots, T$, and  $\tilde{z}(n)$ is expanded as
\begin{align}
\tilde{z}(n)= \left\{
\begin{aligned}
&\tilde{w}(n),     &     {\rm if}\,\,\,B(n)=0,\\
& \eta \tilde{f}_n\tilde{x}(n)+\tilde{w}(n),      &          {\rm if}\,\,\,B(n)=1.  \\
\end{aligned}
\right.
\end{align}

Define
\begin{align}
U=&|\eta|^2P_{x}P_f=(R+1)|\eta|^2P_{s}\sum_{m=0}^{M}|g_m|^2\sum_{k=0}^{K}|f_k|^2,\\
V=&P_w=2(T+1)N_w.
\end{align}
Thus, the test statistic $\Gamma_t$ is subjected to \cite{Wang_2016ITComm, Yang_2018_TWC}
\begin{align}
\Gamma_t\sim
\left\{
\begin{aligned}
&\mathcal{N}\left(V, \frac{V^2}{W}\right),     &     {\rm if}\,\,\,B(n)=0.\\
&\mathcal{N}\left(U+V, \frac{(U+V)^2}{W}\right),      &          {\rm if}\,\,\,B(n)=1.  \\
\end{aligned}
\right.
\end{align}

The ML detector can be made as
\begin{align}
\hat{B}(n)=\mathop {\arg\max} \limits_{B(n)\in \lbrace 0,1\rbrace}{\rm Pr}(\Gamma_t|B(n)),
\end{align}
where ${\rm Pr}(\Gamma_t|B(n))$ is the probability density function (PDF) of $\Gamma_t$ given $B(n)$.

\subsection{Optimal Threshold}

The optimal threshold $T^{{\rm opt}}_h$
of the ML detector
must satisfy
\begin{align}
{\rm Pr}(\Gamma_t|B(n)=0)={\rm Pr}(\Gamma_t|B(n)=1).
\label{eq:Optimal_threshold}
\end{align}
The solution to (\ref{eq:Optimal_threshold}) is
 derived as
\begin{align}
T^{{\rm opt}}_h=\frac{V(U+V)+\sqrt{V^2(U+V)^2+(2+4V/U)\frac{\ln{(1+U/V)}}{W}}}{U+2V}.
\end{align}

Thus, the detection rule is
\begin{align}
\hat{B}(n)= \left\{
\begin{matrix}
0,  & {\rm if}\,\,\,  \Gamma_t<T^{{\rm opt}}_h,\\
1,   &{\rm if}\,\,\,  \Gamma_t>T^{{\rm opt}}_h.
\end{matrix}
\right.
\end{align}

\subsection{BER Performance}
Define $p_0={\rm Pr}(B(n)=1|B(n)=0)$ and $p_1={\rm Pr}(B(n)=0|B(n)=1)$.
The BER of the ML detector is given by
\begin{align}
P_e= & {\rm Pr}(B(n)=0)p_0+{\rm Pr}(B(n)=1)p_1\nonumber\\
   = & \frac{1}{2}\left(p_0+p_1\right).
\end{align}
Given the detection threshold $T^{{\rm opt}}_h$, there is
\begin{align}
P_e=  \frac{1}{2}+\frac{1}{2}Q\left(\frac{T^{{\rm opt}}_h-V}{\sqrt{\frac{V^2}{W}}}\right)-\frac{1}{2}Q\left(\frac{T^{{\rm opt}}_h-U-V}{\sqrt{\frac{(U+V)^2}{W}}}\right),
\end{align}
where
 \begin{align}
Q(x)=\frac{1}{\sqrt{2\pi}}\int_{x}^{\infty}{\rm e}^{-\frac{t^2}{2}}{\rm d}t.
\end{align}

\subsection{Equiprobable Error Threshold}
 In this subsection,
 we discuss the detection threshold that
 can obtain
the same error probability for
$B(n)=0$  and $B(n)=1$, i.e., $p_0=p_1$ \cite{Wang_2016ITComm}.

We can further derive the equation $p_0=p_1$ as
 \begin{align}
 Q\left(\frac{T^{{\rm equ}}_h-V}{\sqrt{\frac{V^2}{W}}}\right)= Q\left(\frac{U+V-T^{{\rm equ}}_h}{\sqrt{\frac{(U+V)^2}{W}}}\right),
 \label{eq:equiprobable_detector}
 \end{align}
 where
 $T^{{\rm equ}}_h$ is the equiprobable error threshold.

Here, we construct the approximation of the Q-function \cite{Lin_1989_approximate}:
\begin{align}
Q(x)\approx \frac{{\rm e}^{-bx-ax^2}}{2}, \,\, a=0.416, \,\, b=0.717.
\end{align}
Utilizing this approximation in (\ref{eq:equiprobable_detector})
and
exerting some mathematical
manipulations, one obtains the detection threshold $T^{{\rm equ}}_h$
 as
\begin{align}
T^{{\rm equ}}_h=\frac{-\varsigma_1+\sqrt{\varsigma^2_1-4\varsigma_0\varsigma_2}}{2\varsigma_0},
\end{align}
where
\begin{align}
\varsigma_0=& \frac{a\sqrt{W}(U^2+2UV)}{V(U+V)},\\
\varsigma_1=& b(U+2V)-2a\sqrt{W}U,\\
\varsigma_2=& -2bV(U+V).
\end{align}

\section{Simulation Results}
\label{sec:Simulation Results}
In this section, numerical results are
provided to assess the BER performance of the ML detector.
All the channels follow Gaussian distributions
with zero-mean and unit-variance.
We set the three parameters $L$, $M$ and $K$ as $L=M=K=8$.
The attenuation  $\eta$ and the noise variance $N_w$
  are fixed as $0.5$ and $1$, separately.
We exert $10^7$
Monte Carlo trials
for every experiment.

\begin{figure}[t]
\centering
\includegraphics[height=55mm,width=92mm]{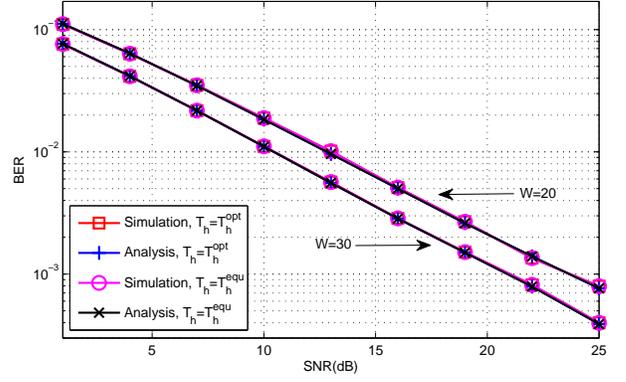}
\caption{BER performance of the proposed design versus SNR with different numbers of averaging samples $W$. }
\label{fig:BER_SNR_simulation} \vspace{-3mm}
\end{figure}

Fig. \ref{fig:BER_SNR_simulation}
plots the BER curves versus signal-to-noise ratio (SNR)
for the  ML detector with two different thresholds.
Two different numbers  of averaging samples $W$, i.e., $W=8$ and $W=10$, are adopted,
and the length of CP  is set to 256.
As seen, the simulation results keep consistent with analysis results.
Besides, the BER performance is enhanced with enlarging SNR.

\begin{figure}[t]
\centering
\includegraphics[height=57mm,width=92mm]{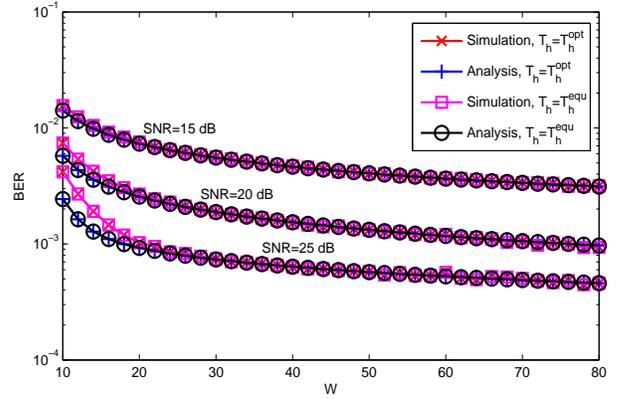}
\caption{BER for the ML detector versus the average number W with different SNR. }
\label{fig:BER_W_simulation} \vspace{-3mm}
\end{figure}

Fig. \ref{fig:BER_W_simulation}
depicts the BER curves versus the number of averaging samples $W$
for three different SNR, i.e., SNR=15 dB, SNR=20 dB and SNR=25 dB.
We fix the length of CP $C$ as 256.
It is found that
 the BER performance is
improved with increasing $W$.

\section{Conclusion}
\label{sec:Conclusion}
This letter investigated the
ambient backscatter communication systems
over frequency-selective channels  and
  proposed an  interference-free     transceiver design
to cope with  the signal detection challenge  at the reader.
The CP structure of OFDM
symbols
was exploited
 to cancel the signal interference at the reader and
meanwhile  our design led to no interference to legacy receivers.
 Finally, a ML detector
 with two thresholds
  was derived
 and its BER expression was obtained.
 Simulation results corroborated that
our transceiver design fitted the scenario of frequency-selective channels
and
 achieved low bit error rate (BER) due to
interference cancellation.

\end{document}